\documentclass[pre,aps,12pt]{revtex4}
\usepackage{amssymb}
\usepackage{graphicx}

\begin{document}

\title{Kagome-like chains with anisotropic ferromagnetic and
antiferromagnetic interactions}
\author{D.~V.~Dmitriev}
\author{V.~Ya.~Krivnov}
\email{krivnov@deom.chph.ras.ru}
\affiliation{Institute of Biochemical Physics of RAS, Kosygin str. 4, 119334, Moscow,
Russia.}
\date{}

\begin{abstract}
We consider a spin-$\frac{1}{2}$ kagome-like chain with competing
ferro- and antiferromagnetic anisotropic exchange interactions.
The ground state phase diagram of this model consists of the
ferromagnetic and ferrimagnetic phases. We study the ground state
and the low-temperature properties on the phase boundary between
these phases. The ground state on this phase boundary is
macroscopically degenerate and consists of localized magnon
states. We calculate the ground state degeneracy and corresponding
residual entropy. The spontaneous magnetization has a jump on the
phase boundary confirming the first-order type of the phase
transition. In the limit of a strong anisotropy the spectrum of
the low-energy excitations has multi-scale structure governing the
peculiar features of the specific heat behavior.
\end{abstract}

\maketitle

\section{Introduction}

\begin{figure}[tbp]
\includegraphics[width=0.6\linewidth,angle=0]{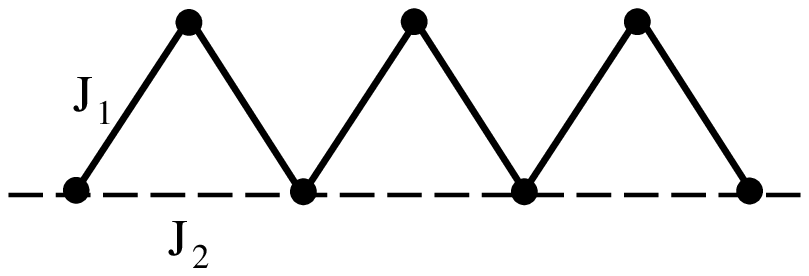}
\caption{The delta-chain model.} \label{Fig_saw}
\end{figure}

The low-dimensional quantum magnets on geometrically frustrated
lattices have been extensively studied during last years
\cite{diep,mila}. There is a broad class of highly frustrated
antiferromagnetic spin systems which supports a completely
dispersionless magnon band (flat band) \cite{flat, shulen, zhit,
mak} so that the excitations in this band are localized states.
The localization of the one-magnon states is a base for the
construction of multi-magnon states, because a state consisting of
independent (non-overlapping) localized magnons is an exact
eigenstate. Such systems include, for example, the delta-chain,
the kagome lattice, kagome-like chains, the Tasaki lattice etc. An
important feature of them is the triangular geometry of
antiferromagnetic bonds. Besides, for these models a special
condition of flat band is required. For example, for the delta
chain (Fig.\ref{Fig_saw}) the relation between interactions must
be $J_{2}=\frac{1}{2}J_{1}$. It was found \cite{flat} that the
ground states of such systems at the saturation magnetic field
consist of independent localized magnons. The ground state and
low-temperature properties for the antiferromagnetic Heisenberg
models with flat band have been actively studied over last
decades. It was shown that flat band physics may lead to new
interesting phenomena such as the residual entropy at the
saturation magnetic field, the zero-temperature magnetization
plateau and the magnetization jump, an extra low-temperature peak
in the specific heat etc.
\cite{schmidt,honecker,zhitomir,Derzhko,zhit}.

Recently it was found that the localized magnon states can be
supported in a certain frustrated spin system with competing
ferro- (F) and antiferromagnetic (AF) interactions. This model is
the $s=\frac{1}{2}$ delta-chain Heisenberg model with the
ferromagnetic $J_{1}<0$ and the antiferromagnetic $J_{2}>0$
interactions (F-AF delta-chain). For $J_{2}=-\frac{J_1}{2}$ the
localized states exist in this model. However, in contrast with
the AF-AF delta-chain the localized magnon states of the F-AF
chain are exact ground states \textit{at zero magnetic field}.
Besides, the ground state manifold contains the special states
with overlapping magnons (localized multimagnon complexes). So,
the ground state degeneracy in this model is even higher than for
AF-AF delta chain. The properties of such F-AF delta chain have
been studied in Ref.\cite{KDNDR,anis}.

\begin{figure}[tbp]
\includegraphics[width=0.6\linewidth,angle=0]{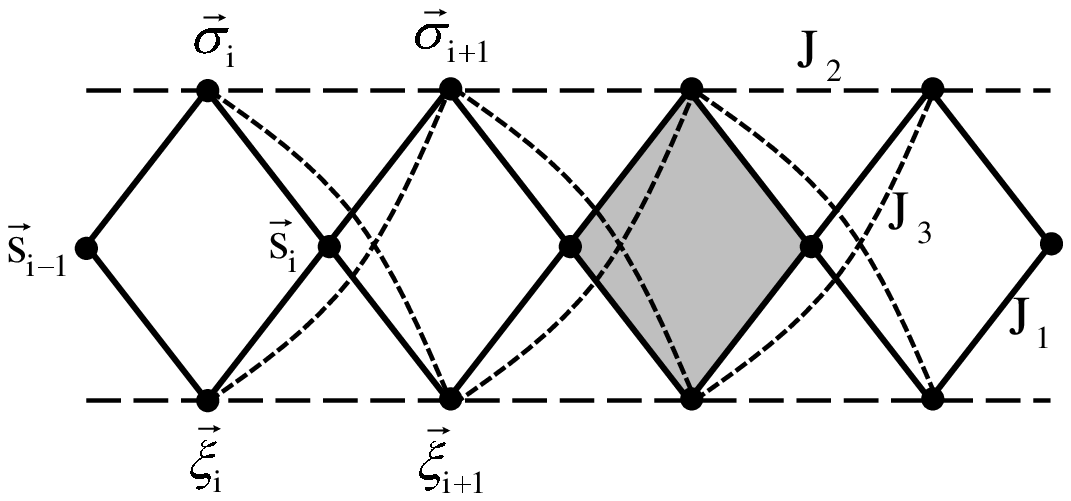}
\caption{The kagome-like spin chain model.} \label{Fig_rhomb}
\end{figure}

In this paper we will study another example of frustrated spin
$s=\frac{1}{2}$ model with ferro- and antiferromagnetic
interactions. It is a kagome-like chain consisting of a linear
chain of corner sharing triangular plaquettes as shown in
Fig.\ref{Fig_rhomb}. The interaction $J_{1}$ between leg and axis
spins is ferromagnetic ($J_{1}<0$). The interaction $J_{2}$ acts
between nearest spins on the same leg, while $J_{3}$ is the
interaction between spins on opposite legs. Both interactions
$J_{2}$ and $J_{3}$ are antiferromagnetic. We will consider two
versions of kagome-like chains, denoted as $(a)$ and $(b)$. In the
model $(a)$ the interaction $J_{3}=0$ and we denote the
interaction $J_{2}$ as $J_{2}=J_{a}$. In the model $(b)$ the
interactions between nearest spins on both the same and the
opposite legs are equal and $J_{2}=J_{3}=J_{b}$. The Hamiltonians
of these F-AF kagome-like chains have the forms:
\begin{equation}
\hat{H}=\sum_{i=1}^n \hat{H}_{i} \label{H}
\end{equation}%
where $\hat{H}_{i}$ is the Hamiltonian of the $i$-th pair of
corner sharing triangles (pair of triangles for brevity) and
$\hat{H}_{i}$ for the kagome-chains $(a)$ and $(b)$ have the forms
\begin{eqnarray}
\hat{H}_{ia} &=&-\sum_{\nu =x,y}(\sigma _{i}^{\nu }+\sigma
_{i+1}^{\nu }+\xi _{i}^{\nu }+\xi _{i+1}^{\nu })s_{i}^{\nu
}+\Delta _{1}[(\sigma _{i}^{z}+\sigma _{i+1}^{z}+\xi _{i}^{z}+\xi
_{i+1}^{z})s_{i}^{z}-1]
\nonumber \\
&&+J_{a}\sum_{\nu =x,y}(\sigma _{i}^{\nu }\sigma _{i+1}^{\nu }+\xi
_{i}^{\nu }\xi_{i+1}^{\nu })+\Delta _{2}(\sigma _{i}^{z}\sigma
_{i+1}^{z}+\xi _{i}^{z}\xi _{i+1}^{z}-\frac{1}{2})  \label{Hia}
\end{eqnarray}%
\begin{eqnarray}
\hat{H}_{ib} &=&-\sum_{\nu =x,y}(\sigma _{i}^{\nu }+\xi _{i}^{\nu
}+\sigma _{i+1}^{\nu }+\xi _{i+1}^{\nu })s_{i}^{\nu }+\Delta
_{1}[(\sigma _{i}^{z}+\xi _{i}^{z}+\sigma _{i+1}^{z}+\xi
_{i+1}^{z})s_{i}^{z}-1]\}
\nonumber \\
&&+J_{b}\sum_{\nu =x,y}(\sigma _{i}^{\nu }+\xi _{i}^{\nu })(\sigma
_{i+1}^{\nu }+\xi _{i+1}^{\nu })+\Delta _{2}[(\sigma _{i}^{z}+\xi
_{i}^{z})(\sigma _{i+1}^{z}+\xi _{i+1}^{z})-1]\}  \label{Hib}
\end{eqnarray}%
where $s_{i}^{\nu}$, $\sigma_{i}^{\nu}$ and $\xi_{i}^{\nu}$ are
$s=\frac{1}{2}$ operators of spins on axis, lower and upper leg
sites, respectively. $J_{a}$ and $J_{b}$ are the antiferromagnetic
leg-leg interactions and we put $J_{1}=-1$. $\Delta_1$ and
$\Delta_2$ are parameters representing the anisotropy of the
axis-leg and the leg-leg exchange interactions respectively, $n$
is the number of axis sites. The periodic boundary conditions
(PBC) are imposed. The constants in Eqs.(\ref{Hia}) and
(\ref{Hib}) are chosen so that the energy of the ferromagnetic
state with the total spin of the system $L^{z}=\pm \frac{3n}{2}$
is zero. As it is seen from Eq.(\ref{Hib}) for the model $(b)$ the
spins on the upper and lower legs present in the Hamiltonian only
in the combinations $(\sigma _{i}^{\nu }+\xi _{i}^{\nu })$ $(\nu
=x,y,z)$, effectively forming the composite spin of pair
$(\vec{\sigma}_{i}+\vec{\xi}_{i})$ which can be either 0 or 1.

The ground state and the low-temperature properties of the
isotropic Heisenberg model on the kagome chain Fig.\ref{Fig_rhomb}
with the antiferromagnetic interactions $J_{1}, J_{2}>0$ and
$J_{3}=0$ has been studied as a function of parameter
$\frac{J_2}{J_1}$ in Ref.\cite{wald}. At
$\frac{J_{2}}{J_{1}}<\frac{1}{2}$ the ground state is
ferrimagnetic of the Lieb-Mattis type \cite{lieb} and at
$\frac{J_2}{J_1}>\frac{1}{2}$ it is the singlet. This model with
$J_2=J_1$ belongs to the class of the flat band models with
multi-magnon states consisting of independent localized magnons.
The ground state and the low-temperature properties of such
AF-kagome-chain with $J_{1}=J_{2}$ have been studied in detail in
Refs.\cite{schnack,honecker}.

In contrast to the antiferromagnetic kagome-chain the models with
the leg-axis ferromagnetic and leg-leg antiferromagnetic
interactions ($J_{1}<0$ and $J_{a}$, $J_{b}>0$) (F-AF
kagome-chains) are less studied. The isotropic kagome-chain $(b)$
with $\Delta_1=\Delta_2=1$ is especially interesting because it is
a minimal model for a description of an interesting class of
quasi-one-dimensional compounds $\mathrm{Ba_3 Cu_3 In_4 O_{12}}$
and $\mathrm{Ba_3 Cu_3 Sc_4 O_{12}}$
\cite{vasil,soos1,soos,vasiliev,indus} and the form of the
Hamiltonian $\hat{H}_{b}$ reflects an unusual topology of these
copper oxides. These oxides represent half-twist ladders in which
successive $\mathrm{CuO}_{4}$ plaquettes are corner shared with
their planes perpendicular to each other. The dominant interaction
in these compounds is between the leg and the axis spins and it is
ferromagnetic \cite{vasil,soos1,soos,vasiliev,indus}. Though the
values of the antiferromagnetic leg-leg interactions are not known
definitely it is expected that they are rather small in comparison
with $J_1$. It means that probably $J_{b}\ll 1$ in these
compounds. Nevertheless, the study of the influence of the
interaction $J_{b}$ on the properties of this model is important
problem. Besides, both F-AF kagome-chains are interesting spin
systems in its own right.

The ground state phase diagram of the isotropic F-AF kagome models
can be analyzed with using of the one-magnon spectrum or the
classical approximation. For example, the minimal energy of
one-magnon excitations over the ferromagnetic state is positive
for $J_{a}<\frac{1}{2}$ ($J_{b}<\frac{1}{4}$) and negative for
$J_{a}>\frac{1}{2}$ ($J_{b}>\frac{1}{4}$). Our numerical
calculations have shown that for $J_{a}>\frac{1}{2}$
($J_{b}>\frac{1}{4}$) the ground state is ferrimagnetic. The
critical points $J_{a}=\frac{1}{2}$ and $J_{b}=\frac{1}{4}$ are
the transition points between these two ground state phases. As
will be shown the F-AF kagome chains have exact localized states
at the transition points $J_{a}=\frac{1}{2}$
($J_{b}=\frac{1}{4}$). However, similarly to the F-AF delta chain
in addition to the multi-magnon configurations consisting of
isolated magnons the special states with overlapping magnons
(localized multi-magnon complex) exist and all of them are exact
ground states \textit{at zero magnetic field}. The ground state
degeneracy in F-AF kagome-chain is macroscopic and higher than for
the antiferromagnetic kagome chain with $J_{2}=J_{1}$. It turns
out that such degeneracy is not exceptional property of the
isotropic F-AF kagome chains and it exists also in the more
general F-AF model with anisotropic exchange interactions for
definite relations between them. The Hamiltonians of such
anisotropic models depend on a single parameter which can be taken
as the anisotropy of the leg-axial interaction $\Delta _{1}$.
These models describe the phase boundary (a transition line)
between different ground state phases on the ($J_{a,b}$, $\Delta
_{1}$) planes: the ferromagnetic phase at $J_{a}<\frac{1}{2\Delta
_{1}}$, $J_{b}<\frac{1}{4\Delta _{1}}$ and the ferrimagnetic phase
at $J_{a}>\frac{1}{2\Delta _{1}}$, $J_{b}>\frac{1}{4\Delta _{1}}$.
The ground state degeneracy on this line does not depend on
$\Delta _{1}$ and grows exponentially with the system size giving
rise to a residual zero-temperature entropy. In the limiting case
$\Delta _{1}=\infty $ the models turn into the Ising models on the
kagome-chains.

The main aim of this paper is to study the F-AF kagome-chains on
the transition line. We will demonstrate that the behavior of the
model on this line has non-trivial peculiarities depending on
$\Delta _{1}$. The limit of the large anisotropy $\Delta _{1}$ is
especially interesting. In this limit the spectrum of low-energy
excitations has a multi-scale structure. This peculiarity of the
spectrum induces the specific properties of the low-temperature
thermodynamics. In particular, the specific heat has many
low-temperature peaks. When the anisotropy $\Delta _{1}$ decreases
the spectrum of excitations is gradually smeared and flattening of
the peaks occurs. In the isotropic case $\Delta _{1}=1$ one
low-temperature maximum in $C(T)$ dependence survives in model
$(a)$ while it transforms to a shoulder in model $(b)$.

In many respects the properties of the kagome-like chains on the
transition line are similar to those for the F-AF delta-chain
studied in Refs.\cite{KDNDR,anis}. Therefore, we will refer to
\cite{KDNDR,anis} for the technical details.

The paper is organized as follows. In section II we study the
ground state properties of the F-AF kagome chains on the
transition line. In section III we study the kagome chains with
large anisotropy $\Delta _{1}$ including the Ising limit. The
structure of the spectrum of low-energy excitations and their
influence on the behavior of the specific heat will be studied.
The properties of the isotropic kagome chains in the transition
point is considered in section IV. In section V we give a summary
of our results.

\section{Ground state degeneracy}

In this section we study the ground state of the F-AF kagome
chains on the phase boundary between the ferromagnetic and
ferrimagnetic phases. At first, we consider the one-magnon states,
i.e. the states in the spin sector $L^{z}=L_{\max }^{z}-1$
($L_{\max }^{z}=\frac{3n}{2}$), where we define the total spin as
$\vec{L}=\sum_{i}(\vec{s}_{i}+\vec{\sigma}_{i}+\vec{\xi}_{i})$ and
denote the quantum number of its magnitude and the z-component by
$L$ and $L^{z}$, respectively. The primitive unit cell of the
kagome chain contains three sites. The corresponding three
branches of one-magnon states with $L^{z}=L_{\max }^{z}-1 $ have
the energies:
\begin{eqnarray}
E_{\pm }(q) &=&\frac{1}{2}\left[ (3\Delta _{1}-J_{a,b}(\Delta _{2}+\cos
q)\pm \sqrt{(\Delta _{1}+J_{a,b}(\Delta _{2}-\cos q)^{2}+4(1+\cos q)}\right]
\nonumber \\
E_{3a}(q) &=& J_{a}\cos q+  \Delta_1-J_{a}\Delta_2,\quad
E_{3b}(q)= \Delta_1 - 2J_{b}\Delta_2
\end{eqnarray}%
where $\alpha =J_{a}$ for model $(a)$ and $\alpha =2J_{b}$ for
model $(b)$.

If we choose exchange interactions as%
\begin{eqnarray}
J_{a} &=&\frac{1}{2\Delta _{1}},\quad \Delta _{2}=2\Delta _{1}^{2}-1
\nonumber \\
J_{b} &=&\frac{1}{4\Delta _{1}},\quad \Delta _{2}=2\Delta _{1}^{2}-1
\label{rel}
\end{eqnarray}%
then the lowest band becomes flat with zero energy, $E_{-}(q)=0$.
For the isotropic model the lowest band is flat if
$J_{a}=\frac{1}{2}$ and $J_{b}=\frac{1}{4}$ for $(a)$ and $(b)$
models, respectively. The energy of the second branch is positive
for all $q$ ($E_{+}(q)\geq 2\Delta _{1}$). The third band for the
model $(a)$ becomes $E_{3a}=(1+\cos q)/2\Delta_1$ and approaches
to zero at $q=\pi$, providing an additional one-magnon state with
zero energy to $n$ states of the lowest band. For the model $(b)$
the third band is flat with the positive energy
$E_{3b}(q)=\frac{1}{2\Delta_1}$.

The dispersionless one-magnon states of lowest band correspond to
localized states which can be chosen for both models $(a)$ and
$(b)$ as
\begin{equation}
\hat{\varphi}_{i}\left\vert F\right\rangle =[s_{i}^{-}+s_{i+1}^{-}+2\Delta
_{1}(\sigma _{i}^{-}+\xi _{i}^{-})]\left\vert F\right\rangle \quad
i=1,\ldots n  \label{5}
\end{equation}%
where $\left\vert F\right\rangle $ is the ferromagnetic state with
all spins up, $s_{i}^{-},\sigma _{i}^{-},\xi _{i}^{-}$ are spin
lowering operators.

In the isotropic case the wave function of the localized magnon is
\begin{equation}
\hat{\varphi}_{i}\left\vert F\right\rangle =[s_{i}^{-}+s_{i+1}^{-}+2(\sigma
_{i}^{-}+\xi _{i}^{-})]\left\vert F\right\rangle \quad i=1,\ldots n
\end{equation}

The wave function $\hat{\varphi}_{i}\left\vert F\right\rangle $ is
localized in a diamond with shaded area in Fig.\ref{Fig_rhomb}. It
can be checked directly that the functions (\ref{5}) are exact
eigenfunction with zero energy of each local Hamiltonian
$\hat{H}_{ia,b}$ (Eqs.(\ref{Hia}),(\ref{Hib})) with exchange
interactions satisfying conditions (\ref{rel}) and, therefore, of
the total Hamiltonian (\ref{H}). All $n$ states (\ref{5}) are
linear independent \cite{KDNDR}. For the model $(a)$ there is in
addition another linear independent exact wave function with zero
energy. It is non-localized one and it has a form
\begin{equation}
\hat{\phi}\left\vert F\right\rangle =\sum_{i=1}^{n}(-1)^{i}(\sigma
_{i}^{-}-\xi _{i}^{-})\left\vert F\right\rangle  \label{wv2}
\end{equation}

Thus, the ground state energy in the spin sector $L^{z}=L_{\max
}^{z}-1$ is zero and there are $n$ or $(n+1)$ such states for the
kagome-chains $(b)$ or $(a)$. In the isotropic case the operator
$\vec{L}$ commutes with the Hamiltonian (\ref{H}) and there are
$(n-1)$ linear combinations of $\hat{\varphi}_{i}\left\vert
F\right\rangle $ which belong to the states with $L=L_{\max }-1$
and one combination belongs to $L=L_{\max }$. The latter is
\begin{equation}
\sum \hat{\varphi}_{i}\left\vert F\right\rangle
=2L_{tot}^{-}\left\vert F\right\rangle .
\end{equation}%
The non-localized wave function (\ref{wv2}) in the model $(a)$
belong to $L=L_{\max }-1$.

The ground state energy in any spin sector can not be negative.
This statement can be proved by a standard way. At first, it can
be checked that 16 eigenvalues of $\hat{H}_{ia}$ and 10
eigenvalues of $\hat{H}_{ib}$ are zero. All other eigenvalues of
$\hat{H}_{ia}$ and $\hat{H}_{ib}$ are positive. Thus, the ground
state energy $E_0$ of $\hat{H}$ satisfies an inequality $E_0\geq
0$. This inequality turns in an equality for the ground states in
the spin sectors $L^{z}=L_{\max }^{z}$ and $L^{z}=L_{\max
}^{z}-1$. The question is: how many states in other spin sectors
have zero energy?

Let us consider the ground state in the spin spin sector
$L^{z}=L_{\max }^{z}-2$ (two-magnon states). It is evident that
the pairs of the independent (isolated) magnons
$\hat{\varphi}_{i}\hat{\varphi}_{j}\left\vert F\right\rangle $
($\left\vert i-j\right\vert >1$) are the eigenfunctions with zero
energy of each local Hamiltonian (and, therefore, of total
Hamiltonian) in this spin sector. However, such states do not
exhaust all possible ones with zero energy. For example, we can
write in the isotropic case the exact two-magnon state as
\begin{equation}
(\hat{\varphi}_{i-1}+\hat{\varphi}_{i}+\hat{\varphi}_{i+1})\hat{\varphi}%
_{i}\left\vert F\right\rangle  \label{b2}
\end{equation}

The wave function (\ref{b2}) is exact one with zero energy because
\begin{equation}
(\hat{\varphi}_{i-1}+\hat{\varphi}_{i}+\hat{\varphi}_{i+1})\hat{\varphi}%
_{i}\left\vert F\right\rangle
=2L_{tot}^{-}\hat{\varphi}_{i}\left\vert
F\right\rangle -\sum_{\left\vert i-j\right\vert >1}\hat{\varphi}_{j}\hat{%
\varphi}_{i}\left\vert F\right\rangle  \label{q}
\end{equation}%
and both function in the right hand of Eq.(\ref{q}) are the exact
wave functions with zero energy. We note that the function
(\ref{b2}) contains overlapping magnons. It can be shown that
pairs of isolated magnons and $n$ eigenfunctions (\ref{b2}) are
linearly independent \cite{KDNDR}. Thus, for the model $(b)$ the
complete manifold of ground states in the sector $L^{z}=L_{\max
}^{z}-2$ consists of $\frac{n(n-3)}{2}$ pairs of isolated magnons
and $n$ eigenfunctions (\ref{b2}) so the ground state degeneracy
in this spin sector is $C_{n}^{2}$. However, for the model $(a)$
there are also two additional exact two-magnon non-localized wave
functions with zero energy. One of them is
\begin{equation}
\hat{\varphi}_{\sigma }\hat{\varphi}_{\xi }\left\vert F\right\rangle
\end{equation}%
where
\begin{equation}
\hat{\varphi}_{\sigma (\xi )}=\sum_{i=1}^{n}(-1)^{i}\sigma ^{-}(\xi ^{-})
\end{equation}%
and another function is $L_{\mathrm{tot}}^{-}\hat{\phi}\left\vert
F\right\rangle $. The ground state degeneracy of the model $(a)$
in the spin sector $L^{z}=L_{\max }^{z}-2$ is $C_{n}^{2}+2$.

The two-magnon wave function of type (\ref{b2}) can be written
also for the anisotropic model with exchange interactions
(\ref{rel}). It has a form
\begin{equation}
\hat{\varphi}_{i}(\hat{\varphi}_{i-1}+B\hat{\varphi}_{i}+\hat{\varphi}%
_{i+1})\left\vert F\right\rangle
\end{equation}%
where $B=2\Delta_1^2-1$ and the degeneracy of the ground state for
the models $(a)$ and $(b)$ is the same as for the isotropic model.

It is evident that eigenfunctions composed of $k$ isolated magnons
are the exact eigenfunction of the ground state in the spin sector
$L^{z}=\frac{3}{2}n-k$. In addition, the construction of the wave
functions of type (\ref{b2}) can be extended for $k>2$. Analysis
similar to that for the F-AF delta-chain \cite{KDNDR} shows that
total number of the exact ground states functions, $G_{n}(k)$, for
fixed $L^{z}=L_{\max }^{z}-k$ is
\begin{eqnarray}
G_{na}(1) &=&C_{n}^{1}+1  \nonumber \\
G_{na}(k) &=&C_{n}^{k}+2,\qquad 2\leq k\leq \frac{n}{2}  \nonumber \\
G_{na}(k) &=&C_{n}^{n/2}+2,\qquad \frac{n}{2}<k\leq \frac{3n}{2}  \label{Ga}
\end{eqnarray}%
\begin{eqnarray}
G_{nb}(k) &=&C_{n}^{k},\qquad 0\leq k\leq \frac{n}{2}  \nonumber \\
G_{nb}(k) &=&C_{n}^{n/2},\qquad \frac{n}{2}<k\leq \frac{3n}{2}  \label{Gb}
\end{eqnarray}

The total degeneracy of the ground state $W$ is
\begin{eqnarray}
W_{a} &=&2^{n}+2nC_{n}^{n/2}+3n+1  \nonumber \\
W_{b} &=&2^{n}+2nC_{n}^{n/2}  \label{W}
\end{eqnarray}

We note that the degeneracy of the ground state given by
Eqs.(\ref{Ga}) and (\ref{Gb}) is valid for any $\Delta_1$ on the
transition line. Eqs.(\ref{Ga}) and (\ref{Gb}) have been confirmed
by an exact diagonalization of finite kagome chains up to $N=24$.

According to Eqs.(\ref{W}) the degeneracy of the ground state of
the kagome-chains on the transition line is exponentially large.
It leads to the residual entropy $s_{0}=\ln (W)/N$ ($N=3n$). In
the thermodynamic limit $N\to \infty $ the residual entropy is the
same for both kagome-chain models and it is equal to
\begin{equation}
s_{0}=\frac{1}{3}\ln 2
\end{equation}

Thus, the entropy per spin is finite at $T=0$. Another unusual
property of the macroscopic degenerate ground state is the
existence of the non-zero spontaneous magnetization  at $T\to 0$.
To show this we calculate the contribution to the partition
function from only the degenerate ground states. Using
Eqs.(\ref{Ga}) and (\ref{Gb}) we represent such truncated
partition function $Z_{gs}$ of the kagome chains in the magnetic
field $h$ in a form \cite{KDNDR}
\begin{equation}
Z_{gs}=2\sum_{k=0}^{n/2}C_{n}^{k}\cosh \left[
\frac{(\frac{3}{2}n-k)h}{T}\right]
+2C_{n}^{n/2}\sum_{k=0}^{n/2}\cosh \left[
\frac{(\frac{3}{2}n-k)h}{T}\right]
\end{equation}

The magnetization is given by $M=T(d\ln Z/dh)$. Using steepest
descent method for calculation of $Z$ we obtain the expression for
the magnetization at $N\gg 1$ in a form
\begin{equation}
\frac{M}{N}=\frac{1+3\exp (h/T)}{6+6\exp (h/T)}  \label{M}
\end{equation}

At $\frac{h}{T}\to 0$ the spontaneous magnetization is
$\frac{M}{N}=\frac{1}{3}$. It means that the ground state is
magnetically ordered.

\section{Kagome chains with high anisotropy}

In this section we study the low-temperature properties of the
anisotropic F-AF kagome-chains on the transition line. At first we
consider the limit of large $\Delta_1$. At $\Delta_1\to \infty$
models (\ref{Hia}) and (\ref{Hib}) reduce to the Ising models on
kagome-chains, the Hamiltonians of which are
\begin{eqnarray}
\hat{H}_{Ia} &=&-\sum_{i}[s_{i}^{z}(\sigma _{i-1}^{z}+\sigma
_{i}^{z}+\xi _{i-1}^{z}+\xi _{i}^{z})-1]+\sum_{i}(\sigma
_{i}^{z}\sigma _{i+1}^{z}+\xi _{i}^{z}\xi
_{i+1}^{z}-\frac{1}{2})-h\sum_{i}(s_{i}^{z}+\sigma _{i}^{z}+\xi
_{i}^{z})  \nonumber \\
\hat{H}_{Ib} &=& \hat{H}_{Ia} - \frac{1}{2}\sum_{i}(\sigma
_{i}^{z}-\xi _{i}^{z})(\sigma _{i+1}^{z}-\xi _{i+1}^{z})
\label{HI}
\end{eqnarray}%
where $h$ is a dimensionless magnetic field and the F and the AF
interactions in $\hat{H}_{Ia}$ are equal in magnitude while in
$\hat{H}_{Ib}$ the AF interaction is half of the F interaction.

The partition function of these Ising models can be obtained using
a transfer-matrix method and it is given by
\begin{equation}
Z=\lambda _{1}^{n}+\lambda _{2}^{n}+\lambda _{3}^{n}+\lambda _{4}^{n}
\end{equation}%
where eigenvalues of the transfer matrix $\lambda _{i}$ for the
model $\hat{H}_{Ia}$ and $\hat{H}_{Ib}$ for $T\ll 1$ are solutions
of equations
\begin{eqnarray}
 \left[
\lambda ^{3}-4\lambda ^{2}\cosh \left( \frac{h}{T}\right) \cosh
\left( \frac{h}{2T}\right) -\lambda - 2\lambda \cosh \left(
\frac{2h}{T}\right) +2\cosh \left( \frac{3h}{2T}\right)
\right]\times
\nonumber \\
\times \left[ \lambda +2\cosh \left( \frac{h}{2T}\right) \right]=0
\end{eqnarray}%
for the model $(a)$ and%
\begin{equation}
\lambda \left[ \lambda ^{3}-2\lambda ^{2}\cosh \left(
\frac{3h}{2T}\right) -4\lambda \cosh \left( \frac{2h}{T}\right)
+4\lambda \cosh ^{2}\left( \frac{h}{2T}\right) +\lambda  -4\cosh
\left( \frac{h}{2T}\right) \right] =0
\end{equation}%
for the model $(b)$.

The eigenvalues $\lambda _{i}$ at $h/T=0$ are
\begin{eqnarray}
\lambda _{1,2a} &=&\frac{5\pm \sqrt{17}}{2},\quad \lambda
_{3a}=-1,\quad \lambda _{4a}=-2
\nonumber \\
\lambda _{1b} &=&4,\quad \lambda _{2b}=\lambda _{3b}=-1,\quad
\lambda _{4b}=0 \label{lambda}
\end{eqnarray}

The ground state of models (\ref{HI}) at $h=0$ has zero energy and
excited states are separated by a gap $\Delta E\simeq 1$. The
ground state is macroscopically degenerate and the total ground
state degeneracy $W$ at $n\gg 1$ is defined by the largest
eigenvalues of (\ref{lambda})
\begin{eqnarray*}
W_{a} &=&(\frac{5+\sqrt{17}}{2})^{n}\simeq 4.56^{n} \\
W_{b} &=&4^{n}
\end{eqnarray*}

We note that the ground state degeneracy is different for the
model $\hat{H}_{Ia}$ and $\hat{H}_{Ib}$.

The ground state degeneracies $G_{In}(k)$ in the spin sector
$L^{z}=(\frac{3n}{2}-k)$ ($0\leq k\leq \frac{3n}{2}$) can be found
as coefficients in the expansion of a partition function $Z/\exp
(\frac{3hn}{2T})$ in powers of $\exp (-h/T)$:
\begin{equation}
Z=\exp \left( \frac{3nh}{2T}\right) \left[ 1+\sum_{k=1}G_{n}(k)\exp \left( -%
\frac{hk}{T}\right) \right]
\end{equation}

In particular,%
\begin{eqnarray}
G_{an}(1) &=&G_{bn}(1)=2C_{n}^{1}  \nonumber \\
G_{an}(2) &=&4C_{n}^{2}-C_{n}^{1},\quad G_{bn}(2)=4C_{n}^{2}-3C_{n}^{1}
\nonumber \\
G_{an}(3) &=&8C_{n}^{3}-4C_{n}^{2}+6C_{n}^{1},\quad
G_{bn}(3)=8C_{n}^{3}-12C_{n}^{2}+14C_{n}^{1}
\end{eqnarray}

To find the magnetization of the Ising model at $\frac{h}{T}\to 0$
we have to evaluate the largest eigenvalue $\lambda _{1}$ up to
the second order in $(\frac{h}{T})^{2}$. Then we have
\begin{eqnarray}
\lambda _{1a} &=&4.56+2.9\left( \frac{h}{T}\right) ^{2}  \nonumber \\
\lambda _{1b} &=&4+2.97\left( \frac{h}{T}\right) ^{2}  \label{lam}
\end{eqnarray}

According to Eqs.(\ref{lam}) the magnetization $M$ $\sim
\frac{h}{T}$ for both Ising models and $M=0$ at $\frac{h}{T}\to
0$. It means that the ground state of the Ising models is
magnetically disordered.

When the anisotropy parameter $\Delta _{1}$ is large but finite it
is convenient to normalize the Hamiltonians on the transition line
as $\hat{H}/\Delta_1$
\begin{equation}
\frac{1}{\Delta _{1}}\hat{H}_{a,b}=\hat{H}_{Ia,b}+\hat{V}_{1a,b}+\hat{V}%
_{2a,b}  \label{Hg}
\end{equation}%
where $\hat{H}_{Ia,b}$ are the Ising Hamiltonians (\ref{HI}) at
$h=0$ and $\hat{V}_{1a,b}$ and $\hat{V}_{2a,b}$ are%
\begin{eqnarray}
\hat{V}_{1a} &=&\hat{V}_{1b}=-2g\sum_{i=1}^{n}\sum_{\nu =x,y}(\sigma
_{i}^{\nu }+\sigma _{i+1}^{\nu }+\xi _{i}^{\nu }+\xi _{i+1}^{\nu
})s_{i}^{\nu }  \nonumber \\
\hat{V}_{2a} &=&2g^{2}\sum_{i=1}^{n}\sum_{\nu =x,y}(\sigma _{i}^{\nu }\sigma
_{i+1}^{\nu }+\xi _{i}^{\nu }\xi _{i+1}^{\nu })-2g^{2}\sum_{i=1}^{n}(\sigma
_{i}^{z}\sigma _{i+1}^{z}+\xi _{i}^{z}\xi _{i+1}^{z}-\frac{1}{2})  \nonumber
\\
\hat{V}_{2b} &=&g^{2}\sum_{i=1}^{n}\sum_{\nu =x,y}(\sigma _{i}^{\nu }+\xi
_{i}^{\nu })(\sigma _{i+1}^{\nu }+\xi _{i+1}^{\nu
})-g^{2}\sum_{i=1}^{n}[(\sigma _{i}^{z}+\xi _{i}^{z})(\sigma _{i+1}^{z}+\xi
_{i+1}^{z})-1]  \label{V}
\end{eqnarray}%
where $g=\frac{1}{2\Delta _{1}}$ is the small parameter.

At $g=0$ the ground state of the Hamiltonians (\ref{HI}) is
$4.56^{n}$ ($4^{n}$)-fold degenerated for the model $(a)$ ($(b)$)
and the other states from the total $8^{n}$ ones are `highly'
excited with $E\geq 1$. The terms $\hat{V}_{1}$ and $\hat{V}_{2}$
lift the degeneracy for each spin sector, but only partly: some
part of the ground state levels remains degenerated with zero
energy while other levels move up. It is remarkable that though
the ground state degeneracies of the limiting Ising models
$\hat{H}_{Ia}$ and $\hat{H}_{Ib}$ are different, the ground state
degeneracies of both models (\ref{Hg}) at any finite value
$\Delta_1$ are the same (excluding minor difference) and they are
given by Eqs.(\ref{Ga}), (\ref{Gb}).

The analysis based on the ED calculations of finite kagome-chains
shows that the spectrum of these removed states is very similar to
that in the strongly anisotropic delta-chain \cite{anis}.
According to the results of Ref.\cite{anis} the spectrum of
low-lying $k$-magnon states (with $k>2$) at $g\ll 1$ consists of
$k$-subsets: the ground states with the energies $E=0$; $k$-magnon
bound complexes with $E\sim g^{2(k-1)}$; the states consisting of
one $(k-1)$-magnon bound complex and one isolated magnon ($E\sim
g^{2(k-2)}$); the states consisting of one $(k-2)$-magnon bound
complex and two isolated magnons ($E\sim g^{2(k-3)}$); and so on.
The highest subset of excitations has the energies $E\sim g^{2}$.
Thus, the low-lying excitations in the sector with
$L^{z}=\frac{N}{2}-k$ are distinctly divided into the parts with
the energies $E\sim g^{2},$ $E\sim g^{4},\ldots E\sim g^{2(k-1)}$.

Taking into account all the states with all the possible values of
$L^{z}$, we found that the total spectrum of the model (\ref{Hg})
can be rank-ordered in powers of the small parameter $g^{2}$ and
that it has a multi-scale structure.

\begin{figure}[tbp]
\includegraphics[width=0.9\linewidth,angle=0]{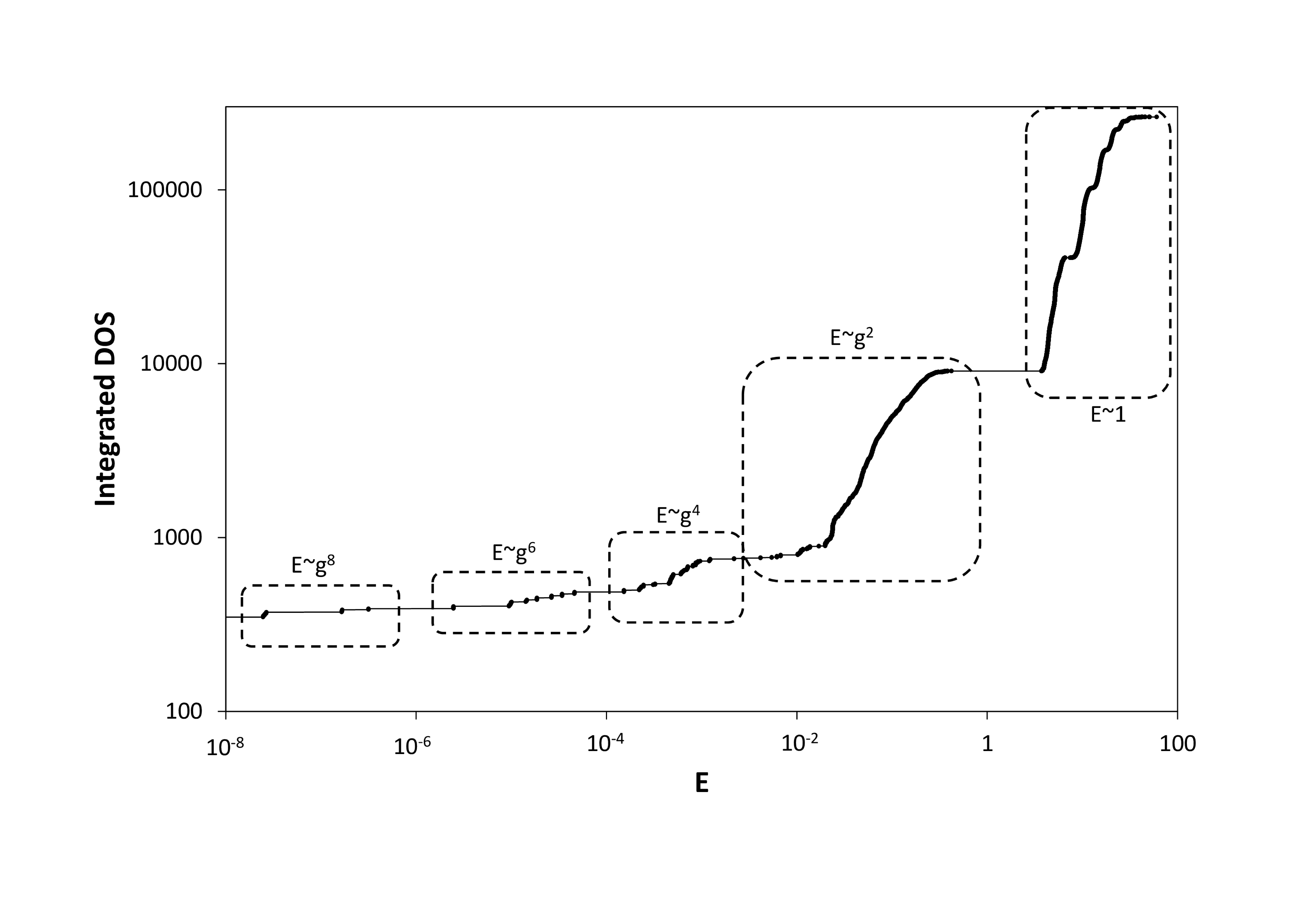}
\caption{Integrated density of states of model $(a)$ for $N=18$
and $\Delta_1=5$.} \label{Fig_DOS1}
\end{figure}
%

The distribution of the energy levels for $N=18$ and $g=0.1$ for
the model $(a)$ is shown in Fig.\ref{Fig_DOS1} (model $(b)$ has a
very similar picture of energy levels). As it can be seen in
Fig.\ref{Fig_DOS1} the spectrum is distinctly divided into some
parts. Each part of the spectrum behaves as $E\sim g^{2k}$ as
depicted in Fig.\ref{Fig_DOS1}. This fact was confirmed
numerically by the comparison of the energies for $g=0.1$ and
$g=0.125$.

\begin{figure}[tbp]
\includegraphics[angle=-90,width=0.9\linewidth,angle=0]{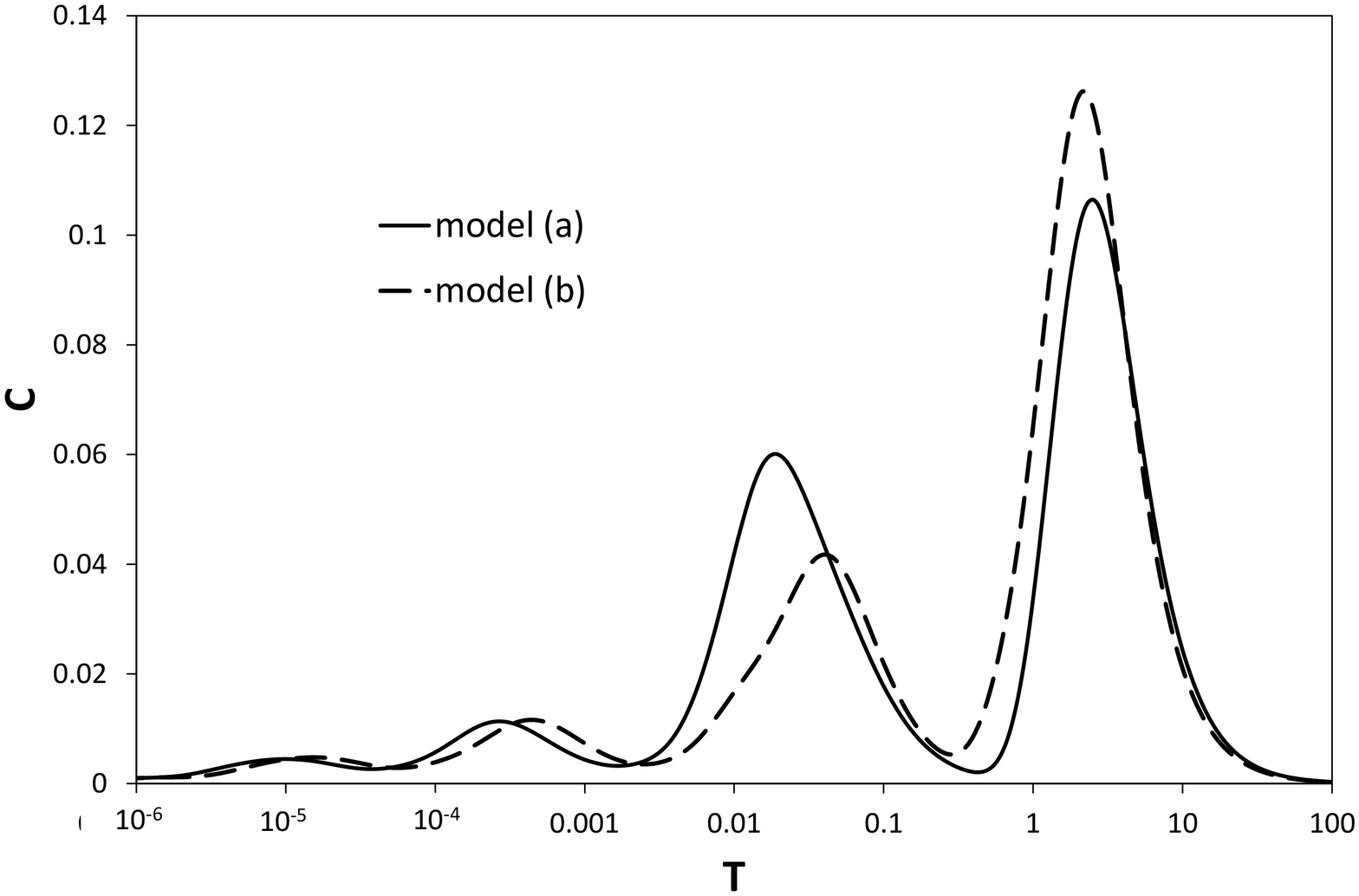}
\caption{Dependence of the specific heat on the temperature for
$N=18$ and $\Delta_1=5$.} \label{Fig_CA10}
\end{figure}

Such structure of the spectrum determines the characteristic
features of the low-temperature thermodynamics. To study the
thermodynamics of model (\ref{Hg}) we use the exact
diagonalization of finite kagome chains with PBC up to $N=18$. In
Fig.\ref{Fig_CA10} we represent the data for the specific heat
$C(T)$ (per site) for $\Delta_1=5$ ($g=0.1$) obtained by the ED
for $N=18$. The temperature dependence of the specific heat shown
in Fig.\ref{Fig_CA10} exhibits numerous maxima for both models
$(a)$ and $(b)$. The peak at $T\geq 1$ is formed by the states
with $E\geq 1$ and the low-temperature maxima are related to the
corresponding parts of the spectrum with the energies $E\sim
g^{4}$, $E\sim g^{6}$ and so on.

\begin{figure}[tbp]
\includegraphics[angle=-90,width=0.9\linewidth,angle=0]{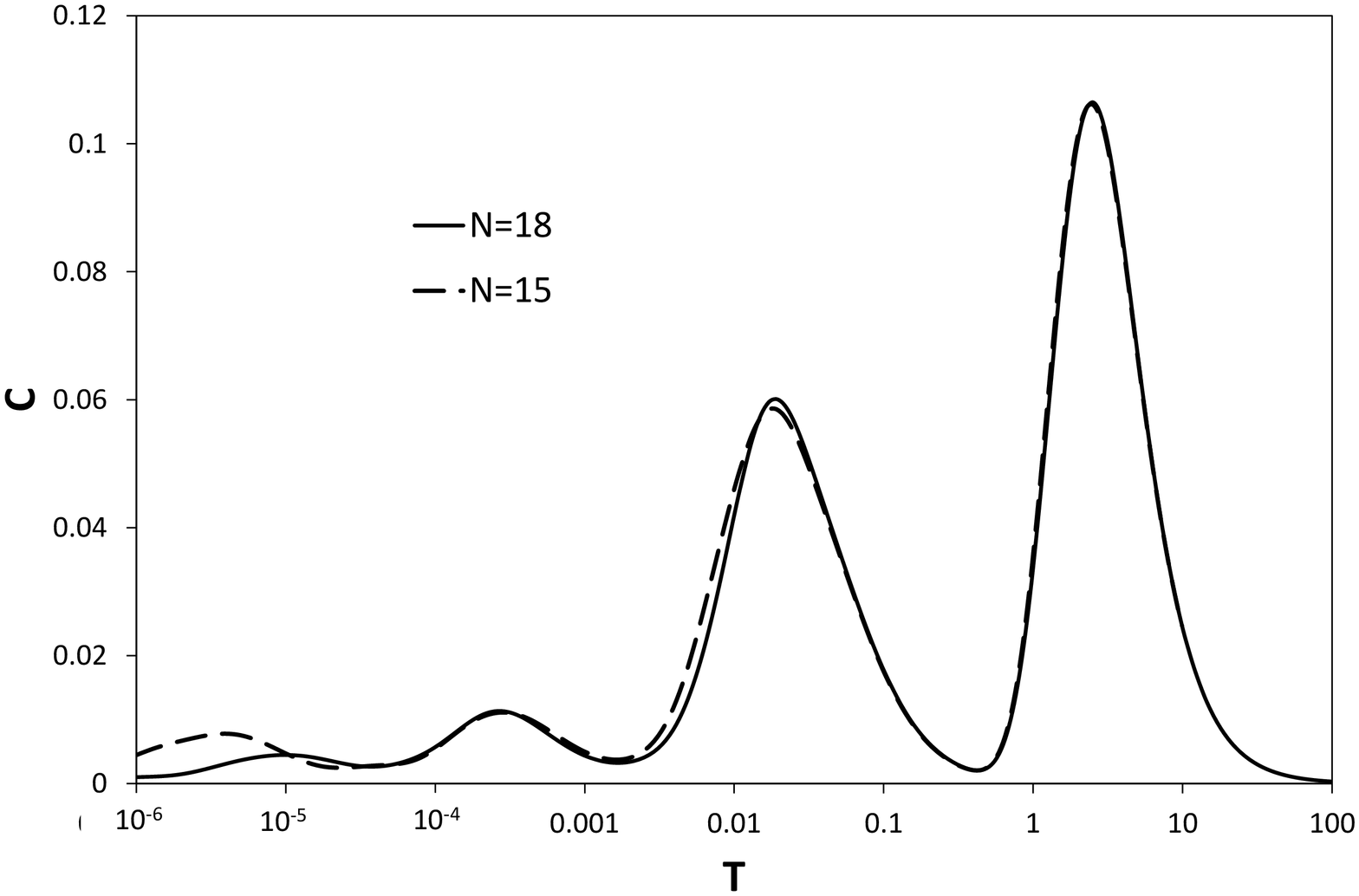}
\caption{Dependence of the specific heat on the temperature for
model $(a)$ at $\Delta_1=5$ and for $N=15,18$.}
\label{Fig_A10N15N18}
\end{figure}

To estimate the finite-size effects we compare the data of $C(T)$
for $\Delta_1=5$ for $N=15$ and $N=18$ (model a). As it can be
seen in Fig.\ref{Fig_A10N15N18} the data of $C(T)$ for $N=15$ and
$N=18$ deviate from each other for $T\lesssim 10^{-4}$ but they
are very close for $T\gtrsim 10^{-4}$. This indicates that the
obtained finite-size data correctly describe the thermodynamic
limit for temperatures $T\gtrsim 10^{-4}$. (We note that a similar
conclusion is valid for other values of the anisotropy). The
deviation of the data for $N=15$ and $N=18$ in the region
$T\lesssim 10^{-4}$ means that the finite-size effects become
essential for $T\lesssim 10^{-4}$ and that the correct description
of the thermodynamics in this temperature region requires the
consideration of larger systems. Nevertheless, the multi-scale
structure of the spectrum for $\Delta _{1}\gg 1$ will lead
certainly to the existence of many maxima in $C(T)$ and their
number is proportional to the system size.

\begin{figure}[tbp]
\includegraphics[angle=-90,width=0.9\linewidth,angle=0]{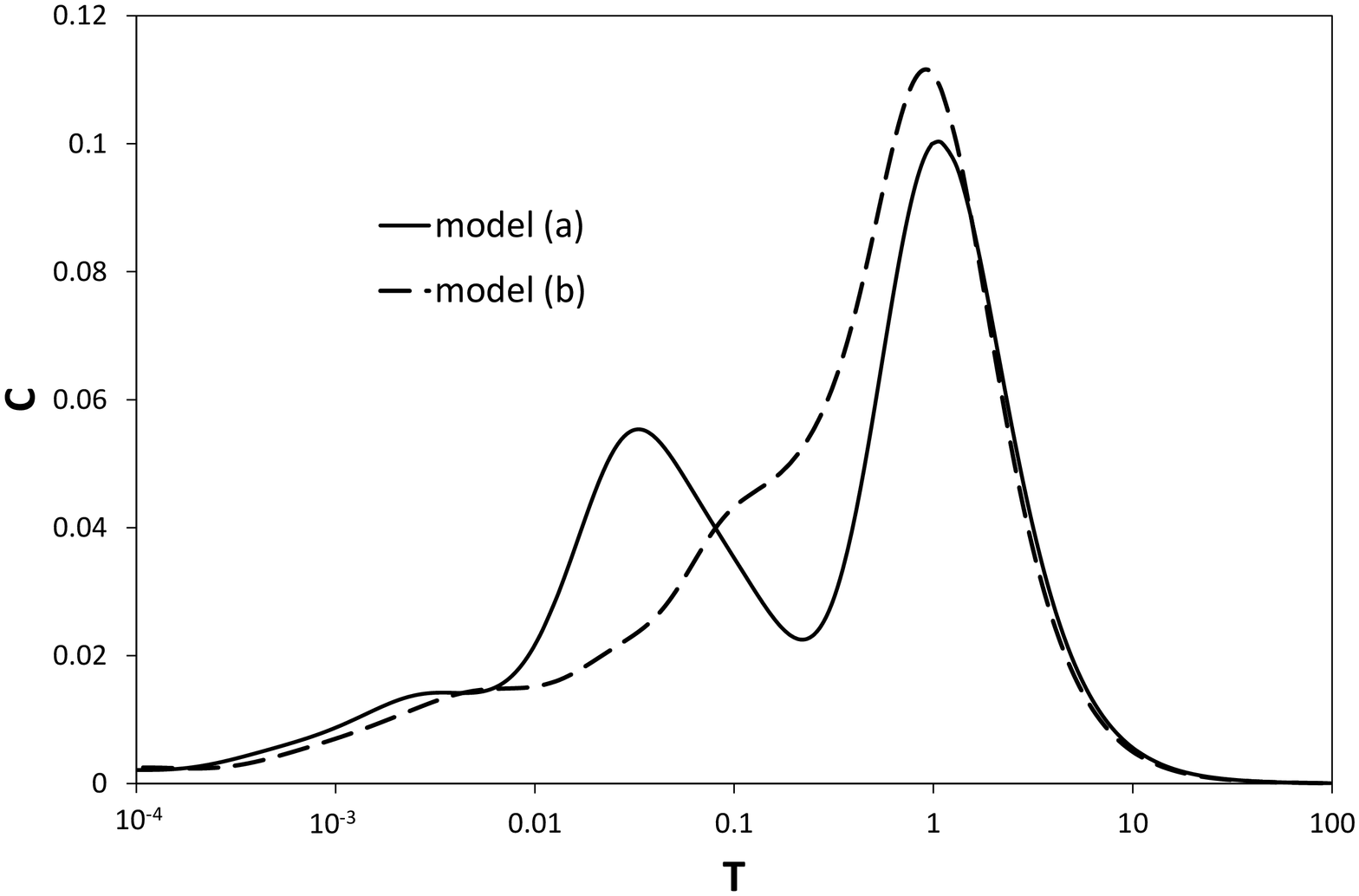}
\caption{Dependence of the specific heat on the temperature for
$N=18$ and $\Delta_1=2$.} \label{Fig_CA4}
\end{figure}

\begin{figure}[tbp]
\includegraphics[angle=-90,width=0.9\linewidth,angle=0]{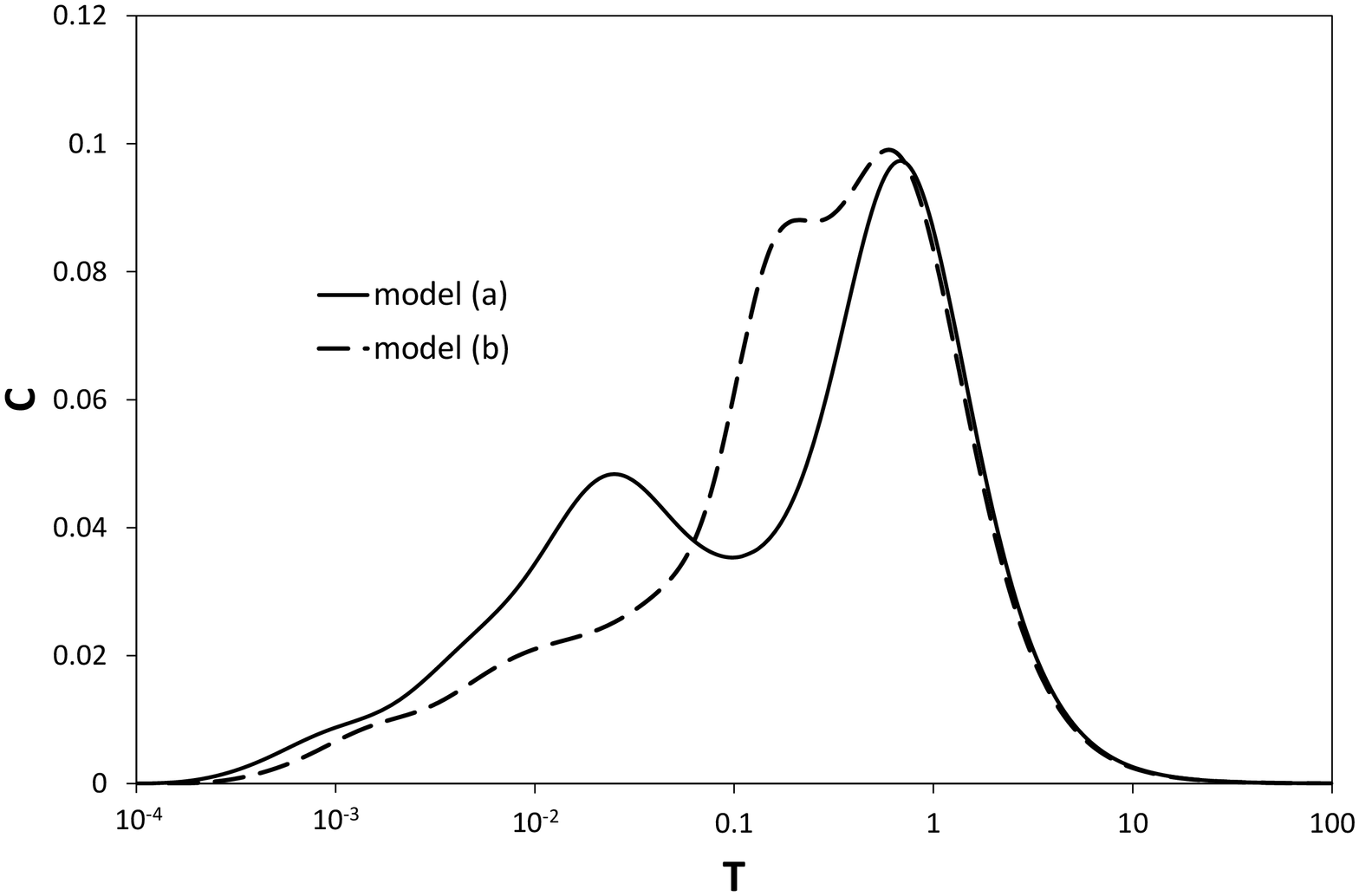}
\caption{Dependence of the specific heat on the temperature for
$N=18$ and $\Delta_1=1$.} \label{Fig_CA2}
\end{figure}

With decreasing of $\Delta_1$ the many-peak dependence of $C(T)$
becomes less pronounced and vanishes for $\Delta_1<3$. For
$\Delta_1<3$ the only one low-temperature maximum in $C(T)$ curve
survives for model $(a)$. For model $(b)$ the low-temperature
maximum transforms into shoulder as shown in Figs.\ref{Fig_CA4},
\ref{Fig_CA2}.

\section{Isotropic models}

As we noted before the ground state degeneracy is the same
throughout the whole transition line including the isotropic point
$\Delta_1=\Delta_2=1$. As to the energy gaps for the excited
states they show a sharp decrease with an increase of the number
of magnons, $k$, similarly to the case of large anisotropy
$\Delta_1\gg 1$. However, the behavior of the gap $E(k,n)$ in the
isotropic model is rather specific. The analysis of the numerical
calculations of finite kagome-chains shows that the behavior of
the gap $E(k,n)$ 
is qualitatively different in the sectors $k<\frac{n}{2}$ and
$k>\frac{n}{2}$. For relatively small number of magnons,
$k<\frac{n}{2}$, the gap rapidly decreases with the increase of
$k$, so that for long chains ($n\gg 1$) it can be approximated as
$E(k,n)\sim \exp (-2\gamma k)$ with $\gamma \simeq 2.80$. This
value of the gap represents the energy of $k$-magnon bound
complex. On the boundary between these two sectors, i.e. for
$k=\frac{n}{2}$, the gap is $E_b\sim \exp (-\gamma n)$. When the
number of magnons exceeds the value $\frac{n}{2}$, the gap ceases
to decrease rapidly and only slightly changes with $k$ and $n$, so
that the gap approximately holds its boundary value $E(k,n)\sim
\exp (-\gamma n)$. Thus, the studied system has a quite unusual
property of exponentially small gaps, and this fact is attributed
to the smallness of the energy of the multi-magnon bound
complexes.

at fixed number of magnons $k<\frac{n}{2}$ the gap behaves as
%
%
Now let us examine the type of the phase transition which occurs
on the studied isotropic transition point. At $J_{a}<\frac{1}{2}$
or $J_{b}<\frac{1}{4}$ the ground state is ferromagnetic. In the
transition points $J_{a}=\frac{1}{2}$ and $J_{b}=\frac{1}{4}$ the
spontaneous magnetization is $\frac{M}{N}=\frac{1}{3}$ according
to Eq.(\ref{M}) (such ground state magnetization is on the entire
transition line). It is interesting to find out what is the ground
state magnetization for $J_{a}>\frac{1}{2}$ or
$J_{b}>\frac{1}{4}$, especially in the vicinity of the transition
points. For this aim we have employed numerical calculations with
the use of both the ED method and the density matrix
renormalization group (DMRG) algorithm. The ED is used for the
consideration of the finite kagome chains up to $N=24$ spins. The
DMRG method allows to explore much larger chains. In our
calculations we considered the chains up to $N=240$ spins. Our
results show that the magnetization is $\frac{M}{N}=\frac{1}{3}$
in the close vicinity of the transition points, at least up to
deviation $0.05$ from the transition points. Local magnetizations
in this ferrimagnetic state are $\langle\sigma_i^z\rangle =
\langle\xi_i^z\rangle\simeq 0.3$, $\langle S_i^z\rangle\simeq
0.4$. The spontaneous magnetization decreases with the increase of
$J_{a,b}$ and tends to $\frac{M}{N}=\frac{1}{6}$ at $J_{a,b}\gg
1$. The question about the behavior of the magnetization in the
region of the intermediate values of $J_{a,b}$ requires further
studies.

The main conclusion of our present examination is that the
magnetization has a jump in the transition points from
$\frac{M}{N}=\frac{1}{2}$ on the ferromagnetic side of the ground
state phase diagram to $\frac{M}{N}=\frac{1}{3}$ in the
ferrimagnetic region, justifying the first-order type of the phase
transition.

\section{Summary}

We have studied the ground state and the low-temperature
properties of two models of the kagome-like chains with the
anisotropic ferromagnetic and the antiferromagnetic exchange
interactions. We focus on the model behavior on the transition
line between the ferromagnetic and the ferrimagnetic ground state
phases. This transition line is parameterized by the anisotropy of
the leg-axis interaction $\Delta _{1}$ which varies from $\Delta
_{1}=1$ (the isotropic case) to $\Delta _{1}\to \infty $ (the
Ising model). On this line the ground state manifold consists of
both the localized multimagnon states and the special multimagnon
complexes. The ground state degeneracy is macroscopic and there is
the residual entropy at $T=0$. In the limiting case $\Delta
_{1}=\infty $ the considered models reduce to Ising kagome-chains.
The ground state degeneracy of these Ising models is huge though
it is different for the kagome chains $(a)$ and $(b)$. For finite
anisotropy $\Delta _{1}$ this degeneracy is partially lifted.
Remarkably, firstly the ground state degeneracy is the same
throughout the whole transition line and secondly it is the same
(with minor difference) for both models in spite of different
degeneracy of corresponding limiting Ising kagome-chains.

We found that the kagome chains on the transition line have finite
spontaneous magnetization, while the ground state of these models
in the Ising limit is magnetically disordered (zero
magnetization). It means that the quantum effects induced by the
perturbations $V_{1}$ and $V_{2}$ (Eqs.(\ref{V})) in the classical
(Ising) model with disordered degenerate ground state lead to
magnetically ordered state, so demonstrating the `order by
disorder' phenomenon.

The characteristic feature of the considered models is the jump of
the spontaneous magnetization in the transition point of the
ground state phase diagram of the isotropic models. Our
examination shows that similar jump takes place on the entire
transition line too.

For $\Delta _{1}\gg 1$ the excitation spectrum has multi-scale
structure and is rank-ordered in powers of small parameter $\Delta
_{1}^{-2}$. The number of sections of the spectrum is equal to the
chain length $n$ and the energy of the levels in the $m$-th
section is $E\sim \Delta _{1}^{-2m+2}$ ($m=1,\ldots n$). The
origin of such exponentially low energy levels is the fact that
the $m$-magnon bound complex in this system has the energy
$E_{m}\sim \Delta _{1}^{-2m+2}$. Each $m$-th section of the
spectrum is responsible for the appearance of $m$-th peak in the
specific heat curve $C(T)$. Thus, the number of the peaks in the
specific heat grows with the length of the chain. Numerical
calculations by the ED of finite chains show that such behavior of
$C(T)$ is qualitatively similar for $\Delta_1>3$ but it is
modified in a definite way for $\Delta_1<3$. In particular, the
specific heat has one low-temperature maximum in model $(a)$ and
the shoulder in $C(T)$ dependence for model $(b)$.

The kagome-chain $(b)$ is related to the copper-oxide compounds
$\mathrm{Ba_3 Cu_3 In_4 O_{12}}$ and $\mathrm{Ba_3 Cu_3 Sc_4
O_{12}}$. Though in this model the interchain interaction leading
to the low-temperature phase transition is not taken into account
it can be considered as the minimal model for the description of
the paramagnetic phase of these systems. In these compounds the
antiferromagnetic leg-leg exchange interaction $J_{b}$ is small
and in the first approximation can be neglected so that the model
becomes pure ferromagnetic one. Such ferromagnetic model has been
studied in Refs.\cite{soos1,soos} and it gives rather adequate
description of the paramagnetic phase of real compounds. Though in
the kagome chain considered by us the AF interaction $J_{b}$ is
not small and corresponds to special model parameter we believe
that this model as well as the model $(a)$ show rather unusual
properties which can be observed in compounds of similar
geometrical structure.

In this paper we have studied two version of the kagome chains
corresponding to special choice of the antiferromagnetic leg-leg
exchange interactions: $J_{2}\neq 0$, $J_{3}=0$ and $J_{2}=J_{3}$.
However, our consideration can be easily extended to the more
general case $J_{2}\neq J_{3}$.

The numerical calculations were carried out with use of the ALPS
libraries \cite{alps}.

\end{document}